\documentclass[aps,pra,superscriptaddress,showpacs,twocolumn]{revtex4-1}
\usepackage{amssymb}
\usepackage{graphicx}
\usepackage{dcolumn}
\usepackage{bm}
\usepackage{amsmath}
\usepackage{braket}

\usepackage[usenames]{color}
\usepackage{textcomp}
\usepackage{float}

\def\be{\begin{equation}}
	\def\ee{\end{equation}}
\def\bea{\begin{eqnarray}}
	\def\eea{\end{eqnarray}}

\usepackage[colorlinks,linkcolor=blue,anchorcolor=blue,citecolor=blue,urlcolor=blue,driverfallback=dvipdfm]{hyperref}

\begin{document}
	
	\title{Quantum phases of $sp^2$-orbital bosonic gases in a hexagonal lattice}
	\author{Pengfei Zhang}
	\affiliation{Department of Physics, National University of Defense Technology, Changsha 410073, P. R. China}
	\author{Hui Tan}
	\affiliation{Department of Physics, National University of Defense Technology, Changsha 410073, P. R. China}
	\author{Jianmin Yuan}
	\affiliation{Department of Physics, Graduate School of China Academy of Engineering Physics, Beijing 100193, P. R. China}
	\affiliation{Department of Physics, National University of Defense Technology, Changsha 410073, P. R. China}
	\author{Yongqiang Li}
	\email{li\_yq@nudt.edu.cn}
	\affiliation{Department of Physics, National University of Defense Technology, Changsha 410073, P. R. China}
	\affiliation{ Hunan Key Laboratory of Extreme Matter and Applications, National University of Defense Technology, Changsha 410073, China}
	\date{\today}
	
	\begin{abstract}
		Orbital degree of freedom plays an important role for understanding quantum many-body phenomena. In this work, we study an experimentally related setup with ultracold bosons loaded into hybridized bands of two-dimensional hexagonal optical lattices. We find that the system supports various quantum many-body phases at zero temperature, including chiral superfluid and chiral Mott insulator by breaking time-reversal symmetry, and time-reversal-even insulating phase, based on dynamical mean-field theory. To explain the time-reversal-even phase, a fourth-order orbital-exchange model is derived to explain the underlying mechanics. To relate to experimental situations, we make band-structure calculations to obtain the Hubbard parameters, and show that these orbital ordering phases persist also in the presence of next-nearest-neighbor hopping.
	\end{abstract}
	
	\maketitle

		\section{introduction}
		Quantum simulation plays an important role for understanding difficult quantum problems in physics~\cite{2007_Lewenstein_AP,2008_Bloch_Dalibard_RMP,2015_Lewenstein_RPP,gross2017quantum}, such as quantum magnetism~\cite{sachdev2008quantum} and topological quantum matter~\cite{haldane2017nobel}. Ultracold gases in optical lattices are one of the most promising and flexible quantum simulators for quantum many-body problems with an unprecedented level of control. Different species or hyperfine states of atoms have been loaded into optical lattices~\cite{PhysRevLett.103.245301,PhysRevLett.105.045303}, which are treated as pseudospin degrees of freedom, and significant efforts have been made to explore magnetic phases in ultracold systems~\cite{mazurenko2017cold,sun2021realization,2022arXiv221213983}. Complex optical lattices, such as triangular~\cite{2009Ultracold,struck2011quantum}, hexagonal~\cite{soltan2011multi,lattice-geometry}, Lieb~\cite{Lieblattice} and kagome lattices~\cite{jo2012ultracold}, trigger even more rich physics, as a result of geometric frustration arising when magnetic interactions between adjacent spins on a lattice are incompatible with the lattice  geometry~\cite{1977Toulouse,semeghini2021probing}.
		
In addition to spin, an alternative approach towards optical lattice simulators is based on orbital degrees of freedom, which provide an opportunity to investigate new orbital physics~\cite{2016_Li_Liu_RPP,lewenstein2011optical}. Here, higher-Bloch bands can be implemented as orbital degrees of freedom, where $p$-orbital systems have been explored extensively both in theories~\cite{PhysRevA.74.013607,PhysRevLett.97.190406,PhysRevB.87.224505, PhysRevLett.97.110405,PhysRevA.106.023315,PhysRevResearch.3.033274,PhysRevLett.121.015303,Saugmann_2020} and experiments~\cite{Hemmerich11,Hemmerich2013,Kock_2016,PhysRevLett.121.265301} in recent years. Various interesting phases have been observed, including chiral superfluid~\cite{Hemmerich11} and sliding phases~\cite{PhysRevLett.121.265301}, where the key element is onsite interactions between atoms for building many-body correlations. Recently, special attention has been paid to the complex-lattice setup, and ultracold $^{87}$Rb atoms have been successfully loaded into the $sp^2$-orbital bands of a hexagonal lattice~\cite{p-band_honecomb,p-topological_2021}. In contrast to the square-lattice case~\cite{Hemmerich11,Hemmerich2013,Saugmann_2020}, a special property of this hexagonal system is that it possesses nearly flat dispersion relations around the $K$ and $M$ points of the first Brillouin zone. Distinct phenomena have been observed experimentally even in the weakly interacting regime, including Potts-nematic superfluid~\cite{p-band_honecomb} and chiral superfluid phases~\cite{p-topological_2021} with bosons condensing at $M$ and $K$ points in the first Brillouin zone, respectively. These experiments indicate that nontrivial underlying mechanics appears for the $sp^2$-orbital system in a hexagonal lattice, where temperature and interaction may play important roles for understanding these quantum phenomena. Another open question is that it is still unclear how orbital textures adapt to the hexagonal-lattice geometry in the strongly interacting regime.
		
Motivated by the experiments~\cite{p-band_honecomb,p-topological_2021,Xu2022}, we study a bosonic system in a two-dimensional (2D) hexagonal lattice with alternating deep and shallow wells, and focus on emergent phenomena from multi-orbital effects and lattice geometries. To explore the physics in the strongly correlated regime, a strong laser is utilized to freeze the motional degree of freedom of atoms in the third direction. By adjusting the sublattice potential imbalance, the $s$-orbital of the shallow wells can be resonance with the $p_{x,y}$-orbitals of the deep sites, realizing a $sp^2$-orbital hybridized system with neglecting all the other orbitals. For a sufficient deep lattice, the system can be described by an extended Bose-Hubbard model. It is expected that various quantum phases appear as a result of the $sp^2$-orbital hybridization in the strongly interacting regime.
		
		To explore the many-body physics of the $sp^2$-orbital system, we utilize a bosonic version of dynamical mean-field theory (BDMFT) applied within the full range from small to large coupling.  With BDMFT, local quantum fluctuations have been taken into account to resolve competing long-range orders.  To explore various exotic magnetic or superfluid phases which break lattice-translational symmetry, we implement real-space BDMFT, where self-energy and Green's function capture inhomogeneous quantum phases with exotic orbital textures. We find that the system supports various quantum many-body phases, including chiral superfluid, chiral Mott insulating, and time-reversal-even insulating phases, based on BDMFT. To explain the underlying mechanics for the time-reversal-even Mott phase, a fourth-order orbital-exchange model is derived. Finally, we make band-structure calculations to obtain the Hubbard parameters with hopping terms up to next-nearest neighbors, and map out the many-body phase diagram, which is more closely related to the experimental situation.
		
		The paper is organized as follows. In Sec. II, we introduce the system and the model studied here, as well as the theoretical approach. In Sec. III, we present a detailed discussion of many-body properties of the system. We conclude in Sec. IV.

		\begin{figure}[t]
		\includegraphics[width=\columnwidth]{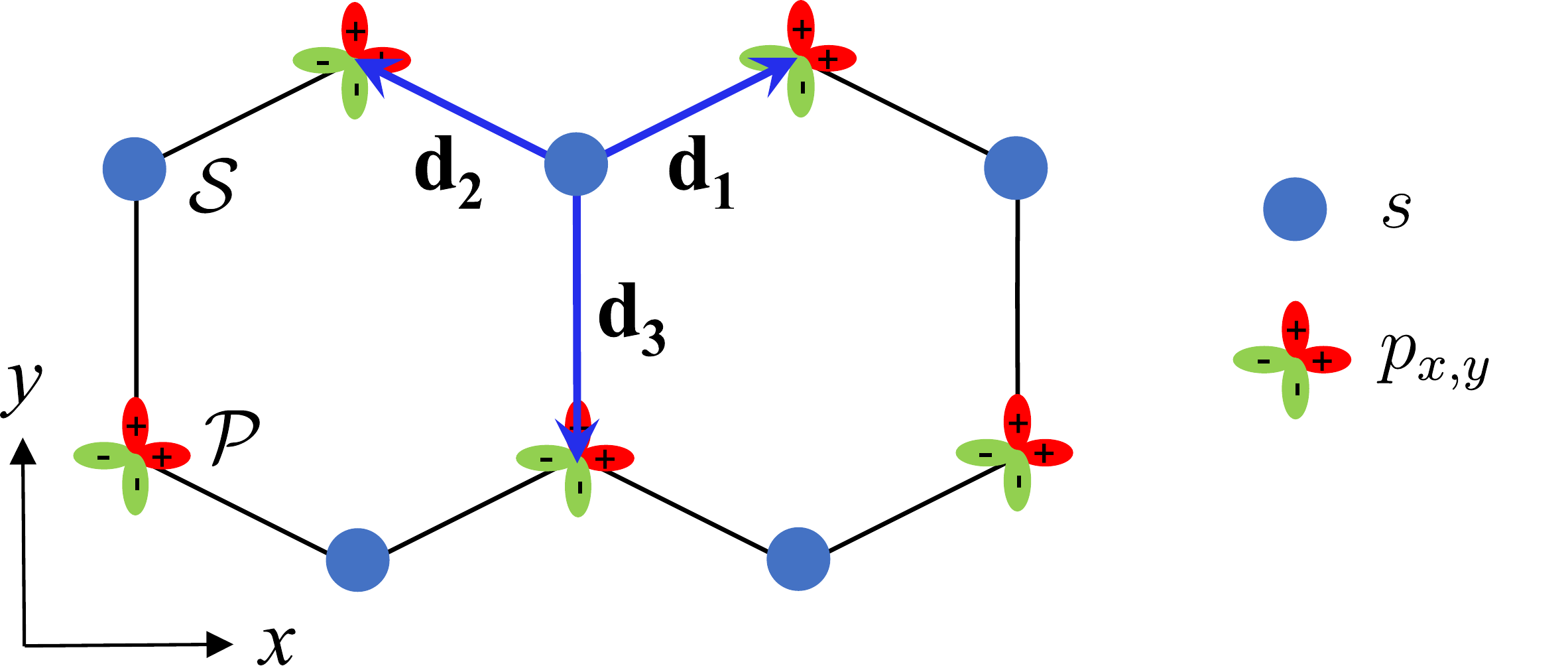}\caption{\label{hex} Setup of the two-dimensional bipartite hexagonal lattice, which possesses two sets of sublattices labeled by $\mathcal{S}$ and  $\mathcal{P}$, respectively. By adjusting sublattice potential imbalance, the $s$- and $p$-orbital bosons can be loaded to the shallow and deep wells, respectively, as achieved in the experiments~\cite{p-band_honecomb,p-topological_2021}, realizing a $sp^2$-orbital hybridized bosonic system in optical lattices.}	
	\end{figure}		
	\section{Model and method}
	\subsection{MODEL}
	We consider a single-component bosonic gas loaded into a hexagonal lattice consisting of two sublattices, denoted as $\mathcal{S}$ and $\mathcal{P}$. By adjusting sublattice potential imbalance~\cite{p-band_honecomb,p-topological_2021}, a $sp^2$-orbital hybridized system can be realized with $\mathcal{S}$ and $\mathcal{P}$ hosting $s$- and $p_{x,y}$-orbitals, respectively, as shown in Fig.~\ref{hex}. The corresponding annihilation operators for the $s$- and $p_{x,y}$-orbital bosonic particles are denoted as $\hat s$, and $\hat p_{x,y}$, respectively. Here, a strong confinement is added to freeze the motional degree of freedom in the third direction, realizing a two-dimensional bipartite lattice system. For a sufficiently deep lattice, the system can be described by a generalized Bose-Hubbard model,
	\begin{eqnarray}  \label{hamiltion}
		\hat{H}&=&t_{s p}\sum_{\mathbf{r} \in \mathcal{S}} \sum_{a=1,2,3 } \left[\hat s_{\mathbf{r}}^{\dagger} (\vec {\hat p}_{\mathbf{r}+ \mathbf{d} _{a}} \cdot \mathbf{e} _{a}) +\text { H.c. }\right] \nonumber \\
		&-&\mu_{s} \sum_{\mathbf{r} \in \mathcal{S}} \ \hat n_{\mathbf{r}, s}-\mu_{p}\sum_{\mathbf{r^{\prime}} \in \mathcal{P} } \left( \hat n_{\mathbf{r^{\prime}}, p_{x}}  +\hat n_{\mathbf{r^{\prime}}, p_{y}}\right) \nonumber \\
		&+&\frac{U_{s}}{2} \sum_{\mathbf{r} \in \mathcal{S}}  \hat n_{\mathbf{r}, s}\left( \hat n_{\mathbf{r}, s}-1\right)+ \sum_{\substack{\mathbf{r^{\prime}} \in \mathcal{P} \\ \sigma=x, y}} \frac{U_{p_{\sigma}}}{2} \hat n_{\mathbf{r^{\prime}}, p_{\sigma}}\left( \hat n_{\mathbf{r^{\prime}}, p_{\sigma}}-1\right) \nonumber \\
		&+& 2 \ U_{p_{xy}} \sum_{\mathbf{r^{\prime}} \in P}  \hat n_{\mathbf{r^{\prime}}, p_{x}}  \hat n_{\mathbf{r^{\prime}}, p_{y}} \nonumber \\
		&+&J  \sum_{\mathbf{r^{\prime}} \in \mathcal{P}}\left(\hat p_{\mathbf{r^{\prime}}, {x}}^{\dagger} \hat p_{\mathbf{r^{\prime}}, {x}}^{\dagger} \hat p_{\mathbf{r^{\prime}}, {y}} \hat p_{\mathbf{r^{\prime}}, {y}}+\text { H.c. }\right),
	\end{eqnarray}
	where the unit vectors $ \mathbf{e} _{1}=\left(\frac{\sqrt{3}}{2}, \frac{1}{2} \right)$, $ \mathbf{e} _{2}=\left(-\frac{\sqrt{3}}{2}, \frac{1}{2} \right)$, $ \mathbf{e} _{3}=\left(0, -1 \right)$, and $\mathbf{d} _{i}= a_{0}\mathbf{e} _{i} $ is the relative position between the two sublattices, with $a_{0}$ being the lattice constant. $t_{s p}$ is the hopping amplitude between the $\mathcal{S}$ and $\mathcal{P}$ sublattices, $\vec {\hat p}_{\mathbf{r}}=\left(\hat p_{\mathbf{r},x}, \hat p_{\mathbf{r},y} \right)$ is the shorthand notation for the annihilation operators $ \hat p_{x}$ and $\hat p_{y}$ at site $\mathbf{r}$, and $\hat n_{\mathbf{r}, \sigma}$ is the number operator for the $\sigma$-orbital at site $\mathbf{r}$. $\mu _{s}$ and $\mu_p$ are the chemical potentials for the $s$- and $p$-orbitals, respectively. $U_{s}$, $U_{p_{x}}$, $U_{p_{y}}$, and $U_{p_{xy}}$ are onsite density-density interactions for $s$-, $p_{x}$-, and $p_{y}$-orbitals, respectively, and $J$ denotes the orbital-changing interaction. According to symmetry analysis, the hexagonal-lattice system possesses $U_{p_{x}}=U_{p_{y}} $ and $J=\frac{U_{p_{x}}-2U_{p_{xy}}}{2}$ for the $p$-orbital interaction terms. In the deep lattice limit, the harmonic approximation yields $U_{p_{x}}=U_{p_{y}}=3U_{p_{xy}}$ \cite{2016Multi}, where the last three terms of Eq.~(\ref{hamiltion}) can be rewritten as
	\begin{equation} \label{p orbital interaction}
		\hat H_{\rm int,\mathcal{P}}=\frac{U_{p_{x}}}{2} \sum _{\mathbf{r} \in \mathcal{P}}  \left( \hat n_{\mathbf{r}, p}^2 - \frac{1}{3}  \hat L_{z,\mathbf{r}}^2\right) ,
	\end{equation}
	with the density $\hat n_{\mathbf{r}, p}=\hat n_{\mathbf{r}, p_{x}}+\hat n_{\mathbf{r}, p_{y}}$, and the orbital angular momentum $ \hat L_{z,\mathbf{r}} = i \left( \hat p_{\mathbf{r}, {x}}^{\dagger} \hat p_{\mathbf{r}, {y}}-\hat p_{\mathbf{r}, {y}}^{\dagger} \hat p_{\mathbf{r}, {x}}\right) $~\cite{PhysRevA.74.013607}.

	\subsection{METHOD}
To understand this generalized Bose-Hubbard model, we utilize BDMFT~\cite{PhysRevB.77.235106,2013Bosonic} to calculate many-body ground states of the system described by Eq.~(\ref{hamiltion}). The advantage of dynamical mean-field theory beyond static mean-field theory is that it includes local quantum fluctuations of the strongly correlated system. The key point of BDMFT is to map the many-body lattice system to a single-site problem, which is then solved self-consistently.  For exploring various exotic magnetic or superfluid phases which break lattice-translational symmetry, we implement a real-space BDMFT (RBDMFT)~\cite{Snoek_2008,PhysRevLett.100.056403,PhysRevLett.105.065301,PhysRevLett.121.093401,PhysRevA.105.063308}. Within RBDMFT, the self-energy is taken to be local, but
depends on the lattice site, i.e. $\sum_{i,j}=\sum_i \delta_{ij}$, where $\delta_{ij}$ is a Kronecker delta. In RBDMFT, our challenge is to solve the single-site problem, and the physics of site $i$ is given by the local effective action $S_{eff}^{i}$, which is given by the standard derivation~\cite{1996Dynamical}. Here, we have two sets of sublattices labeled by $\mathcal{S}$  and   $\mathcal{P}$, which indicates that we need two types of the local effective action $S_{eff}^{\mathcal{S}}$ and $S_{eff}^{\mathcal{P}}$,
\begin{widetext}
	\begin{equation} \label{effective action of S}
		\begin{aligned}
			S_{eff}^{\mathcal{S}}&=\int_{0}^{\beta}  d\tau d\tau^{\prime}
			\sum_{\sigma=x,y}
			\left(\begin{array}{ccc}
				 b_{0,s}^*(\tau)\\
				 b_{0,s}(\tau)
			\end{array}\right) ^T \mathcal{G}_{0, s, s^{\prime},  p_\sigma , p_\sigma^{\prime} }^{-1}(\tau-\tau^\prime)\left(\begin{array}{ccc}
			 b_{0,s^{\prime}}(\tau^{\prime})\\
			 b_{0,s^{\prime}}^*(\tau^{\prime})
			\end{array}\right)  \\
			&+\int_{0}^{\beta}  d\tau \left\lbrace \sum_{\left\langle 0j\right\rangle, \sigma=x,y } t^{ 0j} \left(   b_{0,s}^*(\tau)  \phi_{j,p_\sigma}(\tau) + \phi_{j,p_\sigma}^*(\tau)  b_{0,s}(\tau) \right)  +\frac{U_{s}}{2}  n_{0, s}(\tau)\left(  n_{0, s}(\tau)-1\right)\right\rbrace,
		\end{aligned}
	\end{equation}
\begin{equation} \label{effective action of P}
	\begin{aligned}
		S_{eff}^{\mathcal{P}}&=\int_{0}^{\beta}  d\tau d\tau^{\prime}
		\sum_{\sigma=x,y}
		\left(\begin{array}{ccc}
			 b_{0, p_\sigma}^*(\tau)\\
			 b_{0, p_\sigma}(\tau)
		\end{array}\right) ^T \mathcal{G}_{0,p_\sigma , p_\sigma^{\prime}, s, s^{\prime} }^{-1}(\tau-\tau^\prime)\left(\begin{array}{ccc}
		 b_{0, p_\sigma^{\prime} }(\tau^\prime)\\
		 b_{0, p_\sigma^{\prime} }^*(\tau^\prime)
		\end{array}\right) \\
		& +\int_{0}^{\beta}  d\tau \left\lbrace \sum_{\left\langle 0j\right\rangle , \sigma=x,y } t^{ 0j} \left(   b_{0,p_\sigma}^*(\tau)  \phi_{j,s}(\tau) + \phi_{j,s}^*(\tau)  b_{0,p_\sigma}(\tau) \right) +2 \ U_{p_{xy}}   n_{0, p_{x}}(\tau)   n_{0, p_{y}}(\tau)\right.\\
		& +\left.\sum_{ \sigma=x,y}\frac{U_{p_{\sigma}}}{2}  n_{{0}, p_{\sigma}}(\tau) \left( n_{0, p_{\sigma}}(\tau)-1\right)+J \left( b_{0, {p_{x}}}^*(\tau)  b_{0, {p_{x}}}^*(\tau)  b_{0, {p_{y}}}(\tau)  b_{0, {p_{y}}}(\tau)+\text { H.c. }\right)\right\rbrace ,
	\end{aligned}
\end{equation}
\end{widetext}
where $\quad$ $\quad$ $\quad$ $\quad$ $\quad$ $\quad$ $\quad$ $\quad$ $\quad$ $\quad$ $\quad$ $\quad$ $\quad$ $\quad$ $\quad$ $\quad$ $\quad$ $\quad$ $\quad$
\begin{widetext}
\begin{equation}
	\begin{aligned} \label{local non-interacting propagator}
		&\mathcal{G}_{0,\nu_1,\nu_1^{\prime},\nu_2, \nu_2^{\prime}}^{-1}(\tau-\tau^\prime)=\\
		&\left(\begin{array}{ccc}
			(\partial_{\tau^{\prime}}-\mu_{\nu_1})\delta_{\nu_1,\nu_1^{\prime}}+\sum_{\substack{\left\langle 0j\right\rangle, \left\langle 0j^{\prime}\right\rangle }} t^{0j} t^{0j^{\prime}}G_{j,j^{\prime},\nu_2, \nu_2^{\prime}}^1(\tau,\tau^\prime) & \sum_{\left\langle 0j\right\rangle, \left\langle 0j^{\prime}\right\rangle} t^{0j} t^{0j^{\prime}} G_{j,j^{\prime},\nu_2, \nu_2^{\prime}}^2(\tau,\tau^\prime)\\
			\\
			\sum_{\left\langle 0j\right\rangle, \left\langle 0j^{\prime}\right\rangle} t^{0j} t^{0j^{\prime}}G_{j,j^{\prime},\nu_2, \nu_2^{\prime}}^{2*}(\tau^\prime,\tau) & 	(-\partial_{\tau^{\prime}}-\mu_{\nu_1})\delta_{\nu_1,\nu_1^{\prime}}+\sum_{\left\langle 0j\right\rangle, \left\langle 0j^{\prime}\right\rangle} t^{0j} t^{0j^{\prime}}G_{j,j^{\prime},\nu_2, \nu_2^{\prime}}^1(\tau^\prime,\tau)\\
		\end{array}\right) , \nonumber
	\end{aligned}
\end{equation}
with $ G_{j,j^{\prime},\nu_2, \nu_2^{\prime}}^1(\tau,\tau^\prime) = \left\langle  b_{j,\nu_2}(\tau) b_{j^{\prime},\nu_2^{\prime}}^*(\tau^\prime)\right\rangle _0 - \phi_{j,\nu_2}(\tau)  \phi_{j^{\prime},\nu_2^{\prime}}^*(\tau^\prime) $, and
$G_{j,j^{\prime},\nu_2, \nu_2^{\prime}}^2 (\tau,\tau^\prime)= \left\langle  b_{j,\nu_2}(\tau)   b_{j^{\prime},\nu_2^{\prime}}(\tau^\prime)  \right\rangle_0  -\phi_{j,\nu_2}(\tau)\phi_{j^{\prime},\nu_2^{\prime}}(\tau^\prime)$.
\end{widetext}
Here, $ \mathcal{G}_{0,\nu_1,\nu_1^{\prime},\nu_2, \nu_2^{\prime}}^{-1}$ is a local non-interacting propagator interpreted as a dynamical Weiss mean field which simulates the effects of all other sites. The static bosonic mean-fields are defined as $\phi_{j,\nu}(\tau)=\left\langle  b_{j,\nu}(\tau)\right\rangle_0 $, where $\left\langle ...\right\rangle_0 $ means the expectation value in the cavity system without the impurity site. Actually, it is difficult to resolve this effective action analytically. In order to obtain many-body ground states, we utilize the Hamiltonian representation and express the effective action in terms of the Anderson impurity Hamiltonian ~\cite{PhysRevB.80.245109,PhysRevB.84.144411},
\begin{widetext}
\begin{equation} \label{Anderson impurity Hamiltonian of S}
	\begin{aligned}
		\hat{H}_A^{\mathcal{S}}&=
		\sum_{\left\langle 0j\right\rangle , \sigma} t^{0j}
		\left( \phi_{j,p_\sigma}^* \hat b_{0,s} +\text { H.c. }\right)+\frac{U_{s}}{2} \hat n_{0, s}\left( \hat n_{0, s}-1\right) -\mu_{s}\hat n_{0, s} +\sum_l \epsilon_l \hat a_l^{\dagger} \hat a_l + \sum_{l} \left( V_{l,s} \hat a_l^{\dagger}  \hat b_{0,s}  +W_{l,s}  \hat a_l  \hat b_{0,s}  +\text { H.c. } \right),
	\end{aligned}
\end{equation}
\begin{equation} \label{Anderson impurity Hamiltonian of P}
\begin{aligned}		
	\hat{H}_A^{\mathcal{P}}&=
	\sum_{\left\langle 0j\right\rangle, \sigma } t^{ 0j}
	\left( \phi_{j,s}^* \hat b_{0,p_\sigma} +\text { H.c. }\right)+\sum_{ \sigma} \left[ \frac{U_{p_{\sigma}}}{2} \hat n_{0, p_{\sigma}}\left( \hat n_{0, p_{\sigma}}-1\right) -\mu_{p}\hat n_{0, p_{\sigma}}\right] +2 U_{p_{xy}} \hat n_{0, p_{x}}  \hat n_{0, p_{y}} \\&+J \left(\hat b_{0, {p_{x}}}^{\dagger} \hat b_{0, {p_{x}}}^{\dagger} \hat b_{0, {p_{y}}} \hat b_{0, {p_{y}}}+\text { H.c. }\right)
	+\sum_l \epsilon_l \hat a_l^{\dagger} \hat a_l + \sum_{l,\sigma} \left( V_{l,\sigma} \hat a_l^{\dagger}  \hat b_{0,p_{\sigma}}  +W_{l,\sigma}  \hat a_l  \hat b_{0,p_{\sigma}}  +\text { H.c. } \right),
\end{aligned}
\end{equation}
\end{widetext}
where the bath of condensed bosons is represented by the Gutzwiller term with superfluid order parameter $\phi_{j,\nu}$ for the component $\nu$. The normal bath is described by operators $\hat a_l^{\dagger}$ with energies $\epsilon_l$, where the coupling between the normal bath and impurity site is realized by $V_{l,\sigma}$ and $W_{l,\sigma}$.
By diagonalizing the Anderson Hamiltonian in the Fock basis, the corresponding solution of the impurity model can be obtained, where bath orbitals $n_{bath}=4$ are chosen in our calculations. After diagonalization, we finally obtain the local Green's functions in the Lehmann representation
\begin{equation}	
\begin{aligned}
&G_{ A,\nu \nu^{\prime}}^{1}(i \omega_n)= \\
&\frac{1}{Z} \sum_{m,n} \left\langle m \right| \hat b_{\nu} \left| n \right\rangle  \left\langle n \right| \hat b_{\nu^{\prime}}^{\dagger} \left| m \right\rangle \frac{e^{-\beta E_n }-e^{-\beta E_m }}{E_n - E_m +i\hbar \omega_n} + \beta \phi_\nu \phi_{\nu^{\prime}}^*, \\
&G_{ A,\nu \nu^{\prime}}^{2}(i \omega_n)=\\
&\frac{1}{Z} \sum_{m,n} \left\langle m \right| \hat b_{\nu} \left| n \right\rangle  \left\langle n \right| \hat b_{\nu^{\prime}} \left| m \right\rangle \frac{e^{-\beta E_n }-e^{-\beta E_m }}{E_n - E_m +i\hbar \omega_n} + \beta \phi_\nu \phi_{\nu^{\prime}},
\end{aligned}
\end{equation}
%
where $\omega_n$ denotes Matsubara frequency. Then, the local self energy for each site can be obtained via the Dyson equation:
\begin{equation}
\Sigma_A (i \omega_n)= \mathcal{G}_A^{-1}(i \omega_n) -G_A^{-1}(i \omega_n),
\end{equation}
where $\mathcal{G}_A^{-1}(i \omega_n)$ denotes the non-interacting Weiss Green's function of the Anderson impurity site. In the framework of RBDMFT, we assume that the impurity self-energy $\Sigma_A (i \omega_n)$  coincides with lattice self-energy $\Sigma_{lattice}(i \omega_n)$.  Therefore, we can employ the Dyson equation in real-space representation to
compute the interacting lattice Green's function:
\begin{equation}
\mathbf{G}_{lattice}^{-1}(i \omega_n)= \mathbf{G}_{0}^{-1}(i \omega_n) - \mathbf{\Sigma}_{lattice} (i \omega_n),
\end{equation}
where the non-interacting lattice Green's function $\mathbf{G}_{0}^{-1}(i \omega_n)=( i \omega_n \sigma_z + \bm{\mu})-\mathbf{t}$,  with the matrix of hopping $\mathbf{t}$ determined by lattice structures. Note here that the boldface quantities denote matrices with site-dependent elements. The self-consistency RBDMFT
loop is closed by the Dyson equation to obtain a new local non-interacting propagator.
The new Anderson impurity parameters are then calculated by comparing the old and new Green's functions, and the procedure is then iterated until convergence is reached.

\section{results}

\begin{figure}[htbp]
	
	\includegraphics[width=1\columnwidth]{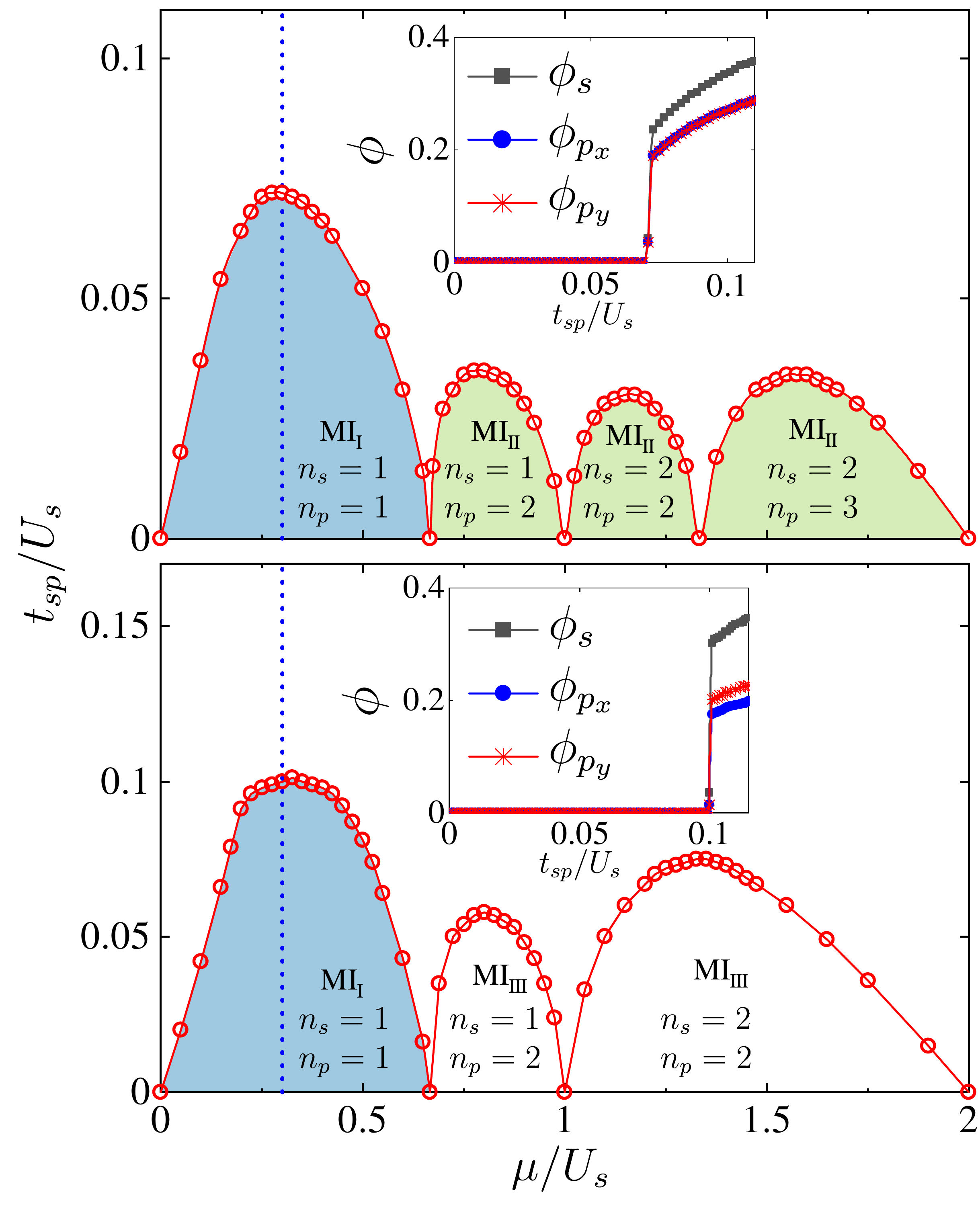}\caption{Phase diagrams of the $sp^2$-orbital hybridized  bosonic atoms in a 2D  bipartite hexagonal lattice. Insets: order parameters $\phi_\nu$ are shown as a function of the hopping amplitude $t_{s p}$ for a fixed chemical potential $\mu /U_{s} =0.3 $ (vertical blue dotted line), indicating a superfluid-Mott insulating phase transition. The chemical potentials are $\mu \equiv \mu_{s}=\mu_{p}$, the interaction strengths $U_{p_{x}}/U_{s}=U_{p_{y}}/U_{s}= 1 $, and $J /U_{s}=1/6$ (upper), and $J /U_{s}=-1/6$ (lower).	 }
\label{phase}	
\end{figure}

\subsection{Many-body phase diagrams}
In the first part, we investigate many-body phase diagrams of the bosonic atoms in a 2D hexagonal lattice for different interactions, based on RBDMFT. To distinguish various quantum phases, the superfluid order parameter is introduced as $\phi_\nu=  \langle \hat b_{\nu} \rangle $, with $\nu= s$, $p_x$, and $p_y$ labeling different orbital states, respectively, and local orbital order $\langle \hat S \rangle = \left[ \langle \hat S^X \rangle, \langle  \hat S^Y\rangle, \langle  \hat S^Z \rangle  \right]  $.  Here, the pseudo-spin operators from the orbital degree of freedom is utilized to quantify orbital order, with $ \hat S_i^X=\frac{1}{2}(\hat b_{i, p_{x}}^{\dagger} \hat b_{i, p_{y}}+ \hat b_{i, p_{y}}^{\dagger} \hat b_{i, p_{x}})$, $ \hat S_i^Y=\frac{1}{2i}( \hat b_{i, p_{x}}^{\dagger} \hat b_{i, p_{y}}- \hat b_{i, p_{y}}^{\dagger} \hat b_{i, p_{x}})$, and $ \hat S_i^Z= \frac{1}{2}(\hat b_{i, p_{x}}^{\dagger} \hat b_{i, p_{x}}- \hat b_{i, p_{y}}^{\dagger} \hat b_{i, p_{y}})$. Accordingly, we define the structure factor of the real-space orbital textures, $ S_{\vec q} =\left| \frac{1}{N_{\rm lat}} \sum_i \langle \hat S_i \rangle e^{i \vec {q} \cdot \vec r_i} \right| $ \cite{2012Bose}, with $N_{\rm lat}$ being the number of lattice sites. To study the $sp^2$-orbital hybridized regime, we first choose a special case with the chemical potentials $\mu_{s}=\mu_{p}\equiv\mu$, and the interaction strengths $U_{p_{x}}=U_{p_{y}}= U_{s}$. To verify finite-size effects, the largest lattice size $N_{\rm lat}=24\times24\times2$ is chosen in our simulations.

Fig.~\ref{phase} displays the many-body phase diagrams for different orbital-changing interactions $J /U_{s}=1/6$ (upper panel), and $J /U_{s}=-1/6$ (lower panel). As expected, the system favors a superfluid phase for larger hopping, and Mott states develop in the lower hopping regime. As shown in the inset of Fig.~\ref{phase}, we clearly observe a first-order Mott-superfluid phase transition. Another typical feature of the many-body phase diagram is the unusual sequence of lower Mott lobes~\cite{Bloch,PhysRevA.101.063611}, as a result of the multi-flavor orbital degrees of freedom. We observe that the phase boundaries for different Mott states are not in the same positions for different sublattices, since the interaction forms of the $\mathcal{S}$ and the $\mathcal{P}$ sites are distinct from each other.
Note here that the case of bipartite square lattice was also discussed \cite{2012Multi}.
\begin{figure}[t]
	\includegraphics[width=\columnwidth]{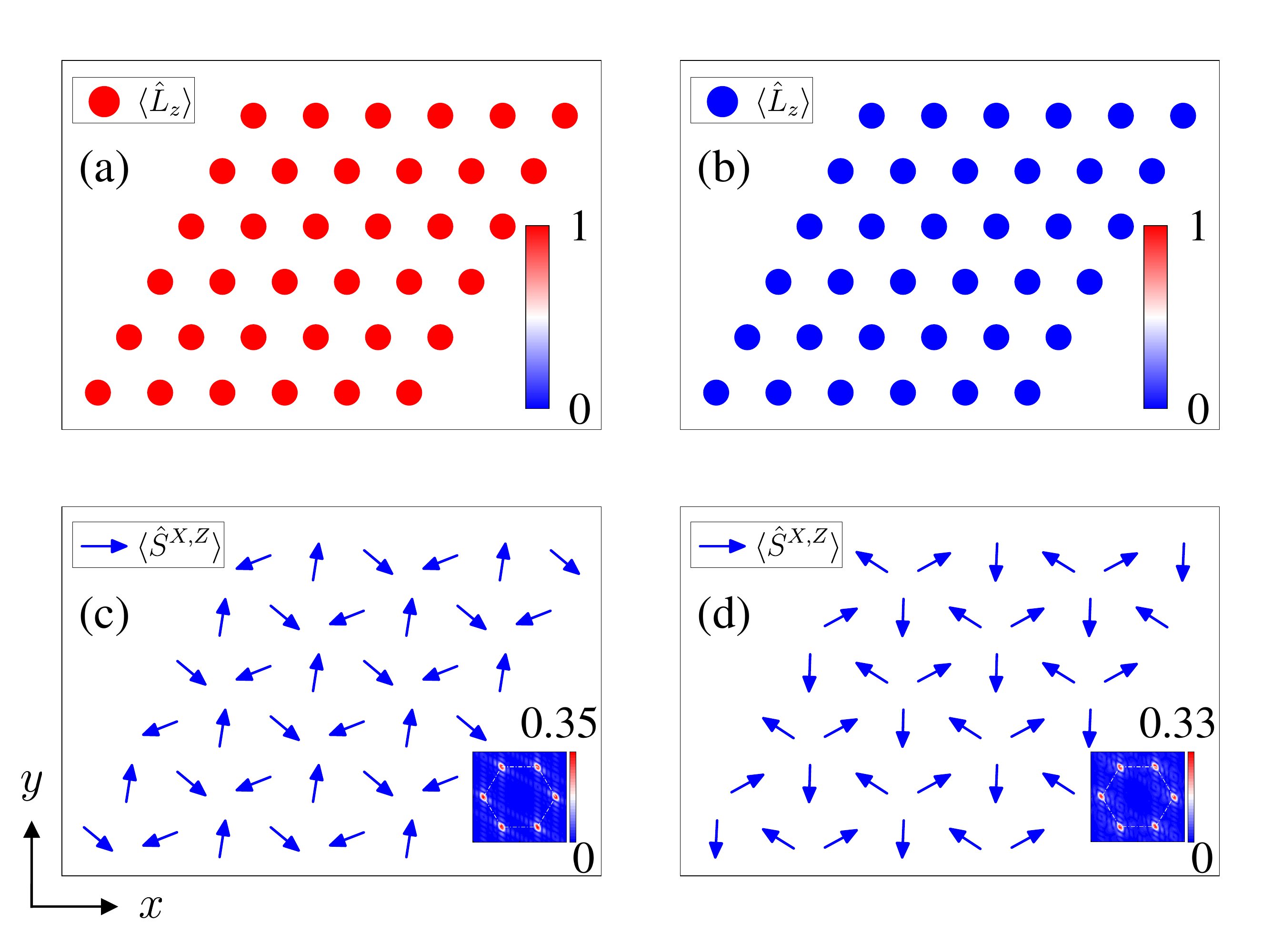}
	\caption{Real-space distributions of orbital textures for the $\mathcal{P}$ sites with $J/U_{s}=1/6$ (a)(c), and $J/U_{s}=-1/6$ (b)(d). (a)(b) real-space distributions of orbital angular momentum $\langle \hat L_z\rangle $ for the superfluid with $\mu=0.3$, and $t_{s p}=0.11$. The color  of the dots represents the value of $\langle \hat L_z\rangle $. (c)(d) real-space orbital textures for the Mott phases with filling $n=1$ for $\mu=0.3$, and $t_{s p}=0.04$, where the blue arrows represent real-space distributions of local orbital order $\langle \hat S^{X,Z}\rangle $ for the $\mathcal{P}$ sites. Insets: Contour plots of static orbital order structure factor $ S_{\vec q}$.  }
\label{orbit-order}
\end{figure}

\begin{figure*}[th!]
	\includegraphics[width=2\columnwidth]{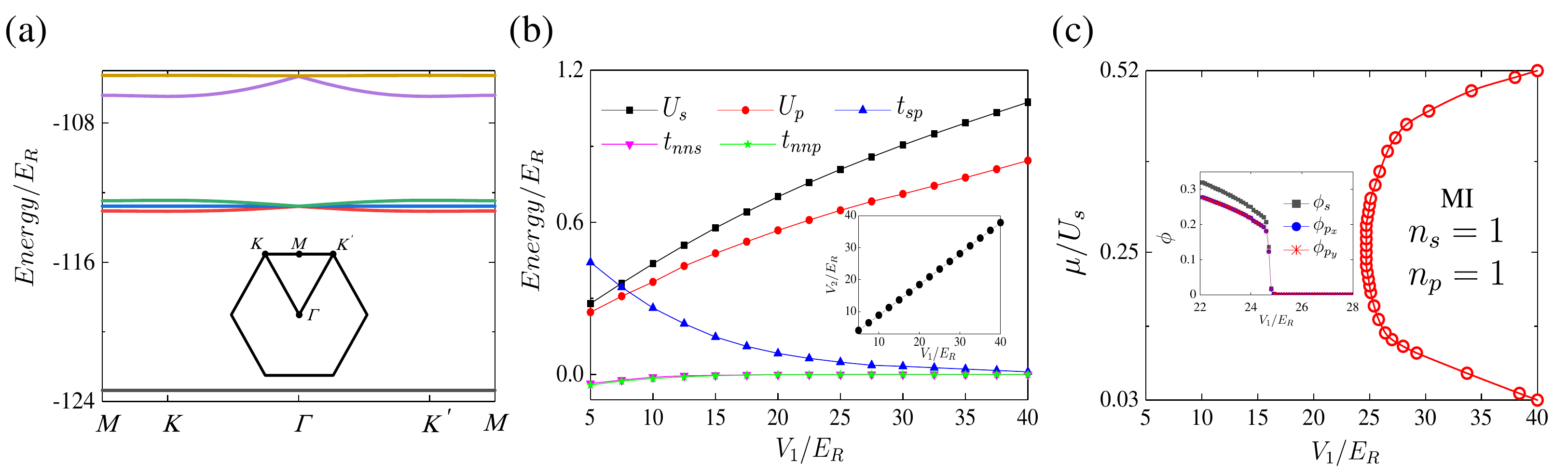}
	\caption{ \label{band structure and phase} (a) Band structure of $V_1=15E_R$ and $V_2=13.6252E_R$, which have triple band crossings between the 2nd, 3rd, and 4th bands, indicating the $s$-orbital being in resonance with the $p_{x,y}$-orbitals, where $E_R$ is the recoil energy. (b) Interaction and  hopping parameters as a function of lattice depth $V_1$. Inset: $s$- and $p_{x,y}$-orbitals resonance for different lattice depths $V_1$ and $V_2$. (c) Phase diagram of the $sp^2$-orbital hybridized bosonic atoms in a 2D hexagonal lattice, where the Hubbard parameters are obtained from band-structure simulations. Inset: order parameters $\phi_\nu$ are shown as a function of $V_1$ for a fixed chemical potential $\mu /U_{s} =0.25 $, indicating the superfluid-Mott-insulating phase transition.}	
\end{figure*}

RDMFT also resolves long-range orbital order of the many-body phases, since it takes higher-order orbital fluctuations into account in the simulations.  We observe an orbital-changing-interaction dependent orbital orders. For positive interaction with $J /U_{s}=1/6$, we find nonzero orbital angular momentum $\langle \hat L_z\rangle\neq0$ both in the superfluid and Mott phases ($\rm MI_{II}$ with $n>1$) by breaking time-reversal symmetry. As shown in Fig.~\ref{orbit-order}(a), real-space orbital texture of the $\mathcal{P}$ sites demonstrates a homogeneous orbital angular momentum $\langle \hat L_z\rangle$ for the superfluid, where the atoms condense in the $K$ point of the first Brillouin zone [inset of Fig.~\ref{band structure and phase}(a)], consistently with experimental observations~\cite{p-topological_2021}. The nonzero value of angular momentum in the phases is not surprising, since the $p$-orbital interaction terms, which are described by Eq.~(\ref{p orbital interaction}), favor the angular momentum order. For negative interaction $J /U_{s}=-1/6$, however, it is expected that $\langle \hat L_z\rangle=0$  both in the superfluid [Fig.~\ref{orbit-order}(b)] and Mott phases ($\rm MI_{I}$ and $\rm MI_{III}$) [Fig.~\ref{orbit-order}(d)] to lower the energy of the system. But we do not observe Potts-nematic condensing at the $M$ point for negative orbital-changing interactions, which was experimentally observed recently~\cite{p-band_honecomb}, and the physical reason may be due to the single-site solver used in our BDMFT approach.

\begin{figure}[t] 	
	\includegraphics[width=\columnwidth]{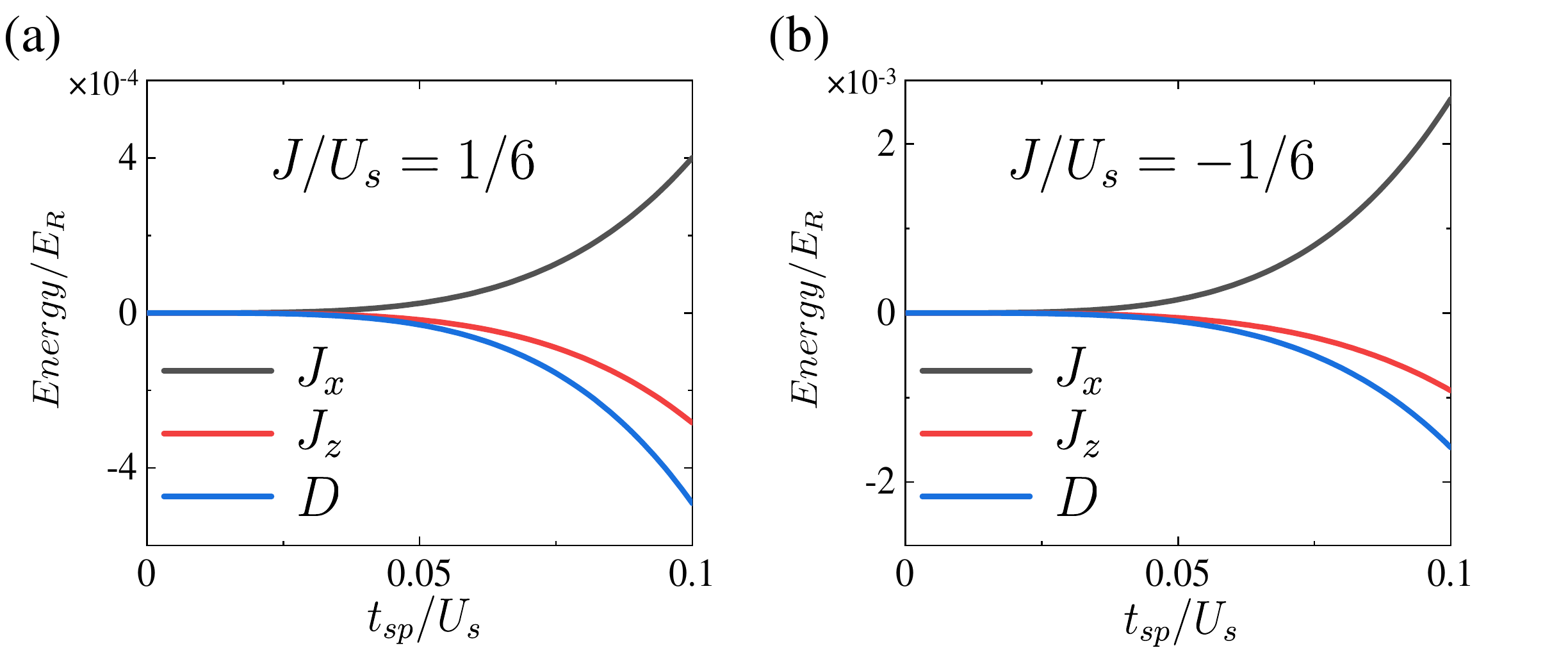}
	\caption{ Coupling strengths of the orbital-exchange model as a function of the tunneling amplitudes for $J/U_{s}=1/6$ (a), and $J/U_{s}=-1/6$ (b). $J_{x}/J_{z}=-3$ and $D/J_{z}=\sqrt{3}$ for arbitrary hopping, as a result of rotational symmetry of the hexagonal lattice.
	}
\label{coupling strengths}
\end{figure}

Interestingly, we find a $120^{\circ}$ in-plane orbital order in the Mott phase ($\rm MI_{I}$) with filling $n=1$ both for positive $J /U_{s}=1/6$ [Fig.~\ref{orbit-order}(c)] and negative interactions $J /U_{s}=-1/6$ [Fig.~\ref{orbit-order}(d)], where $3\times 3 $ orbital textures of $\langle \hat S^{X,Z}\rangle $ appear with $\langle \hat L_z\rangle=0$ by respecting time-reversal symmetry. To understand the physical phenomena in the Mott-insulating phase with filling $n=1$, we need an effective orbital-exchange model for the deep Mott regime. The effective orbital-exchange Hamiltonian is obtained by considering the tunneling part as a perturbation to the full Hamiltonian~\cite{PhysRevLett.91.090402,essle_2005,Mila2011,PhysRevLett.111.205302}. In the strong coupling limit $t_{sp}\ll U_s$, we can use two projection operators $\hat{P}$ and $\hat{Q}=1-\hat{P}$ to divide the Hilbert space into two orthogonal subspaces. Here, $\hat{P}$ projects onto the subspace $\mathcal{H}_P$ with only one atom occupied per site, and $\hat Q$ projects onto the complementary subspace $\mathcal{H}_Q$ (See the Appendix for details).
For Hamiltonian $\hat{H}$, we divide it into two parts $\hat{H}$ = $\hat{H}_t +\hat{H}_U $, that $\hat{H}_t$ describes tunneling terms and $\hat{H}_U$ describes interaction terms. The Schr\"{o}dinger equation reads

\begin{equation} \label{sh equation}
	\hat{H}\left| \psi \right\rangle =\left( \hat{H}_t +\hat{H}_U \right)\left( \hat P +\hat Q \right) \left| \psi \right\rangle = E\left| \psi \right\rangle,
\end{equation}
which leads to an effective Hamiltonian $H_{eff}$ in the Mott phase with unit filling
\begin{equation} \label{H_eff}
	\hat{H}_{eff} =  \hat P  \hat{H}_t \hat Q  \frac{1}{E-\hat Q \hat{H}\hat Q } \hat Q  \hat{H}_t \hat P .
\end{equation}
Because $E\sim t^2/U$
, we obtain $\frac{1}{E-\hat Q \hat{H}\hat Q } \approx \frac{1}{-\hat Q \hat{H}_U\hat Q -\hat Q \hat{H}_t\hat Q}$. Using $\frac{1}{A-B}=\frac{1}{A} \sum_{n=0}^{\infty}(B\frac{1}{A})^n$,  with $A = -\hat Q \hat{H}_U\hat Q $ and $B = \hat Q \hat{H}_t\hat Q $, we obtain the effective Hamiltonian
%
\begin{equation} \label{Heff}
\hat{H}_{eff} =  \hat P  \hat{H}_t \hat Q \frac{1}{-\hat Q \hat{H}_U\hat Q } \sum_{n=0}^{\infty} \left( \hat Q \hat{H}_t\hat Q  \frac{1}{-\hat Q \hat{H}_U\hat Q } \right)^n \hat Q  \hat{H}_t \hat P.
\end{equation}
Since the system have two sets of sublattices, the second-order terms are then trivial. Take a $p$-orbital atom as an example, it can tunnel to its neighboring $\mathcal{S}$ site as a $s$-orbital atom, and then the $s$-orbital atom can only tunnel back to the empty $\mathcal{P}$ site as a $p$-orbital atom. This is nothing but an onsite energy shift. Thus, it is not possible to generate effective interaction terms between the orbitals via second-order processes. Therefore,  one must include fourth-order terms $\mathcal{O}\left( t^4 / U^3 \right)$, which give rise to the coupling between $\mathcal{P}$ sites to reach a nontrivial effective Hamiltonian. The effective orbital-exchange model is finally given by:
\begin{equation} \label{effective hamiltion}
	\hat H_{eff}=\sum_{\left\langle ij \right\rangle } J_{x}\hat S_i^X \hat S_j^X+J_{z}\hat S_i^Z \hat S_j^Z+D\left( \hat S_i \times \hat S_j\right)_y,
\end{equation}
where $\left\langle ij \right\rangle$ denotes the nearest-neighbor sites $i$ and $j$ of the $\mathcal{P}$ sublattice, and the Heisenberg exchange coupling terms $J_{x,z}$ and the Dzyaloshinskii-Moriya term $D$~\cite{1958A,Moriy} are given in the Appendix.
We find the disappearance of the Heisenberg exchange term $J_y$ and the appearance of the Dzyaloshinskii-Moriya interaction in the $y$ direction in the effective exchange model, which is the underlying physics of the disappearance of orbital angular momentum with $\langle \hat L_z\rangle=0$ for the Mott phase with filling $n=1$. Instead, the Dzyaloshinskii-Moriya term competes with the normal exchange terms, and can induce coplanar orbital textures.

In Fig.~\ref{coupling strengths}, the coupling strengths are shown as a function of the tunneling amplitudes. Interestingly, we observe $J_{x}/J_{z}=-3$ and $D/J_{z}=\sqrt{3}$ for arbitrary hopping, as a result of rotational symmetry of the hexagonal lattice. Indeed, the interplay of the Dzyaloshinskii-Moriya and the normal exchange terms results in a $120^{\circ}$-orbital coplanar order for the Mott-insulating phase with $n=1$, as shown in Fig.~\ref{orbit-order}(c)(d), where the blue arrows represent the real-space distribution of local orbital order $\langle \hat S^{X,Z}\rangle $ of the $\mathcal{P}$ sites. This phase is also characterized by the structure factor $ S_{\vec q} $. As  shown in the inset of Fig.~\ref{orbit-order}(c)(d), the structure factor exhibits six peaks at $K$  and $K^\prime $ points.


\subsection{Band-structure simulations and many-body phases}
In the previous part, we study the $sp^2$-orbital hybridized system with ideal Hubbard parameters. In this part, we investigate the robustness of quantum phases against Hubbard parameters, which can be obtained from band-structure simulations. In particular, we consider a two-dimensional bipartite hexagonal lattice potential
	\begin{equation} \label{ potential of hex }
	\begin{aligned}
		V_{\rm hex}(\mathbf{r})=&-V_1\sum_{\substack{\alpha=1,-1 \\ \sigma=1,2,3}}\left[ 3+ e^{i \alpha \left( \mathbf{b_\sigma}\cdot \mathbf{r} - \frac{2\pi}{3}\right)} \right] \\ &-V_2\sum_{\substack{\alpha=1,-1 \\ \sigma=1,2,3}}\left[ 3+ e^{i \alpha \left( \mathbf{b_\sigma}\cdot \mathbf{r} + \frac{2\pi}{3}\right)} \right],
	\end{aligned}
\end{equation}
where $V_1$ and $V_2$ are the lattice depths of the two sets of lattices, $\mathbf{b_1}$ and $\mathbf{b_2}$ are reciprocal lattice vectors for the two-dimensional hexagonal lattice in the $xy$ plane. In the third direction, we consider a strong laser field to freeze the motional degree of freedom with $V_z=50\, E_R$, where $E_R$ is the recoil energy. We choose $\mathbf{b_1}=\frac{4\pi}{\lambda}(\frac{3}{4},-\frac{\sqrt{3}}{4})$ ,  $\mathbf{b_2}=\frac{4\pi}{\lambda}(0,\frac{\sqrt{3}}{2})$ and $\mathbf{b_3}=\mathbf{b_1}+\mathbf{b_2}$  to generate a two-dimensional hexagonal lattice as shown in  Fig.~(\ref{hex}).

Experimentally, the potential difference between the $\mathcal{S}$ and $\mathcal{P}$ wells can be readily adjusted by tuning the ratio $V_1/V_2$, just as already done in the experiments~\cite{p-band_honecomb,p-topological_2021,Xu2022}. In our case, we consider only three bands, i.e. 2nd, 3rd, and 4th bands, which can be isolated from other bands with atoms loading into these bands via band swapping technique~\cite{Kock_2016,p-band_honecomb,p-topological_2021}. Fig.~\ref{band structure and phase}(a) shows the energy spectra of the lowest six-energy bands for $V_1=15\, E_R $ and $ V_2=13.6252\,E_R$, based on a plane-wave expansion. Here, the 2nd, 3rd, and 4th bands are isolated from other bands, and the corresponding orbitals are the $s$-orbital in the shallow $\mathcal{S}$ sites, and two $p$-orbitals in the deeper $\mathcal{P}$ sites, realizing a $sp^2$-orbital hybridized system in a two-dimensional optical lattice.

For a sufficient deep lattice, a tight-binding model can be utilized to describe the system, as shown in Eq.~(\ref{hamiltion}), based on the Wannier-function basis. The corresponding Hubbard parameters, such as interaction and hopping parameters, can be calculated using numerical methods. Here, we calculated the parameters of the Hubbard model using the maximally localized Wannier functions for composite bands~\cite{PhysRevB.56.12847,2011Maximally,2012Tight}, based on the software package~\cite{2013Ab}. In addition, we introduce the next-nearest-neighbor hopping terms $t_{nns}$ and $t_{nnp}$, which are the nearest-neighbor hopping amplitudes within the same sublattice. Under the resonance of the $s$- and $p_{x,y}$-orbitals by controlling the ratio $V_1/V_2$, the hopping amplitudes and interactions as a function of $V_1$ are shown in Fig.~\ref{band structure and phase}(b). Here, we take $^{87}{\rm Rb}$ as an example, and choose the wavelength $\lambda=1064\, {\rm nm}$, and $s$-wave scattering length $a_s=100.4\,a_0$ with $a_0$ being Bohr radius. We find that the next-nearest-neighbor hopping terms decrease quickly, approaching tiny values even for a moderate lattice depth.

Based on the Hubbard parameters obtained from band-structure simulations, we calculate the phase diagram of the $sp^2$-orbital hybridized bosonic system in a two-dimensional hexagonal lattice. Generally, the next-nearest-neighbor hopping terms between $p_{x,y}$-orbitals prefer a Potts-nematic superfluid with $\langle \hat L_z \rangle = 0$ by condensing atoms at the $M$ point [inset of Fig.~\ref{band structure and phase}(a)] of the first Brillouin zone for the hexagonal lattice. However, the next-nearest-neighbor hopping is strongly suppressed for a moderate deep lattice, and, even for the lattice depth $V_1=5\, E_R$, the physics is dominated by the nearest-neighbor hopping by developing chiral superfluid with $\langle \hat L_z \rangle \neq 0$. Upon increasing the lattice depth, the atoms localize, and a Mott insulator develops with a $120^{\circ}$-orbital coplanar order, as shown in Fig.~\ref{band structure and phase}(c) with filling $n=1$.

\section{conclusion}
In summary, we study an experimentally related setup with ultracold bosons loaded into the $sp^2$-orbital hybridized bands of two-dimensional hexagonal optical lattices, and obtain zero-temperature quantum phases, based on bosonic dynamical mean-field theory. A rich phase diagram, including chiral superfluid, chiral Mott insulating, and time-reversal-even insulating phases, is found. In the strongly interacting regime, a fourth-order orbital-exchange model is derived, and a consistent description is found. To relate to experimental observations, we make band-structure calculations to obtain the Hubbard parameters, and resolve various quantum many-body phases, indicating the chance to observe these phases using current experimental techniques.

\begin{acknowledgments}
We acknowledge helpful discussions with Xiaopeng Li, Xiaoji Zhou, Zhifang Xu, and Xu-Chen Yang. This work is supported by the National Natural Science Foundation of China (Grants No. 12074431, and 11774428), Excellent Youth Foundation of Hunan Scientific Committee under Grant No. 2021JJ10044, and NSAF No. U1930403. We acknowledge the Beijing Super Cloud Computing Center (BSCC) and ChinaHPC for providing HPC resources that have contributed to the research results reported within this paper.
\end{acknowledgments}

\newpage
\begin{widetext}
	\section{Appendix}
	\renewcommand{\theequation}{S\arabic{equation}}
	\renewcommand{\thefigure}{S\arabic{figure}}
	\renewcommand{\bibnumfmt}[1]{[#1]}
	\renewcommand{\citenumfont}[1]{#1}
	\setcounter{equation}{0}
	\setcounter{figure}{0}

\subsection{Effective orbital-exchange model}	
The fourth-order orbital-exchange model is given by:

	\begin{equation} \label{fourth-order}
		\hat{H}_{eff} =  \hat P  \hat{H}_t \hat Q \frac{1}{-\hat Q \hat{H}_U\hat Q }  \hat Q \hat{H}_t\hat Q  \frac{1}{-\hat Q \hat{H}_U\hat Q }\hat Q \hat{H}_t\hat Q  \frac{1}{-\hat Q \hat{H}_U\hat Q }  \hat Q  \hat{H}_t \hat P.
	\end{equation}
In the tight-binding regime, we consider a three-site ($ \mathcal{P},\mathcal{S},\mathcal{P}$) problem, and then the subspace $\mathcal{H}_P$, where all lattice sites are occupied with one atom, is
\begin{equation} \label{H_P}
	\mathcal{H}_P:\left\lbrace \ket{p_x,s,p_x},\ket{p_x,s,p_y},\ket{p_y,s,p_x},\ket{p_y,s,p_y}\right\rbrace,
\end{equation}
where $\ket{p_\sigma,s,p_{\sigma^\prime}}$ denotes the orbital state $p_x$ or $p_y$ in the  $\mathcal{P}$ site and $s$ in the $\mathcal{S}$ site. The subspace $\mathcal{H}_Q$, where one lattice site is occupied with two atoms, is
\begin{equation} \label{H_Q}
	\begin{aligned}
	\mathcal{H}_Q:
	&\left\lbrace \ket{0,ss,p_x},\ket{0,ss,p_y},\ket{p_x,ss,0},\ket{p_y,ss,0},\ket{0,s,p_xp_x},\ket{0,s,p_xp_y},\ket{0,s,p_yp_y},\ket{p_xp_x,s,0},\ket{p_xp_y,s,0}, \right.\\
	 &\left.\ket{p_yp_y,s,0},\ket{p_x,0,p_xp_x},\ket{p_x,0,p_xp_y},\ket{p_x,0,p_yp_y},\ket{p_y,0,p_xp_x},\ket{p_y,0,p_xp_y},\ket{p_y,0,p_yp_y},\ket{p_xp_x,0,p_x},\right.\\
	&\left.\ket{p_xp_y,0,p_x},\ket{p_yp_y,0,p_x},\ket{p_xp_x,0,p_y},\ket{p_xp_y,0,p_y},\ket{p_yp_y,0,p_y}\right\rbrace. 		
\end{aligned}
\end{equation}
From these two subspaces, we can obtain the matrix form of $\hat P  \hat{H}_t \hat Q$, $\hat Q \hat{H}_U\hat Q$ and $\hat Q \hat{H}_t\hat Q$. Eq.~(\ref{fourth-order}) yields the effective orbital-exchange model, which is described by Eq.~(\ref{effective hamiltion}).
The three coupling strengths are given by
\begin{equation}
\begin{aligned}
 J_{x}&=\frac{6t_{s p}^4}{2U_{s}^2U_{p_{xy}}}+\frac{6t_{s p}^4}{4U_{s}U_{p_{xy}}^2}+\frac{3t_{s p}^4}{16U_{p_{xy}}^3}+\frac{24Jt_{s p}^4}{-U_{s}^2\tilde{U}}+\frac{24Jt_{s p}^4\left(U_{p_{x}}+U_{p_{y}}\right) }{-U_{s}\tilde{U}^2}+\frac{6t_{s p}^4\left(4J^3+JU_{p_{x}}^2+JU_{p_{y}}^2+JU_{p_{x}}U_{p_{y}} \right) }{-\tilde{U}^3} , \\  J_{z}&=\frac{2t_{s p}^4}{2U_{s}^2U_{p_{xy}}}+\frac{2t_{s p}^4}{4U_{s}U_{p_{xy}}^2}+\frac{t_{s p}^4}{16U_{p_{xy}}^3}+\frac{t_{s p}^4\left(6U_{p_{y}} -2U_{p_{x}}\right) }{-U_{s}^2\tilde{U}}+\frac{t_{s p}^4\left(6U_{p_{y}}^2-2U_{p_{x}}^2+16J^2 \right) }{-U_{s}\tilde{U}^2}\\
 &+\frac{t_{s p}^4\left((3U_{p_{y}}^3-U_{p_{x}}^3)/2+2J^2U_{p_{x}}+10J^2U_{p_{y}} \right) }{-\tilde{U}^3} ,\\
 D&=\frac{\sqrt{3} t_{s p}^4\left(4J+U_{p_{x}}-3U_{p_{y}} \right) }{-U_{s}^2\tilde{U}}+\frac{\sqrt{3} t_{s p}^4\left( 4JU_{p_{x}}+4JU_{p_{y}}+U_{p_{x}}^2-3U_{p_{y}}^2-8J^2\right) }{-U_{s}\tilde{U}^2}\\
 &+\frac{\sqrt{3}t_{sp}^4\left(4J^3+JU_{p_{x}}^2+JU_{p_{y}}^2-J^2U_{p_{x}}-5J^2U_{p_{y}}+JU_{p_{x}}U_{p_{y}}+(U_{p_{x}}^3-3U_{p_{y}}^3)/4 \right) }{-\tilde{U}^3},
\end{aligned}
\end{equation}
with $\tilde{U}=U_{p_{x}}U_{p_{y}}-4J^2$.

\end{widetext}

\bibliography{reference}

\end{document}